\documentclass[12pt]{article}
\usepackage[scaled]{helvet}
\usepackage[T1]{fontenc}
\usepackage{fancyhdr}
\usepackage[sort&compress]{natbib}
\usepackage{ulem}
\usepackage[final]{pdfpages}
\usepackage{psfrag}
\usepackage{graphics}
\usepackage{epsfig}
\usepackage{wrapfig}
\usepackage{shadow}
\usepackage{color}
\usepackage{dcolumn}	
\usepackage{amsmath}
\usepackage{amsfonts}
\usepackage{amssymb}
\usepackage{tcolorbox}
\usepackage{pdfpages}
\usepackage{indentfirst}
\usepackage{multicol}

\usepackage{dcolumn}
\usepackage{bm}

\usepackage[tableposition=top,font={sf}]{caption}
\usepackage{framed}
\usepackage{subcaption}
\usepackage{microtype}
\usepackage[hypertexnames=false]{hyperref}
\usepackage[capitalise,noabbrev]{cleveref}

\newcommand{\shorttitle}{AADT, Richard K. Barry}
\definecolor{grey}{gray}{0.7}
\fancypagestyle{simple}{ \lhead[]{} \rhead[]{} }
\begin{document}
\thispagestyle{simple}

\begin{center} 
{\LARGE \bf Advanced Astrophysics Discovery Technology\\ \vspace{0.1in} in the Era of Data Driven Astronomy} \\[4ex] 

\begin{figure}[!b]
    \centering
    \includegraphics[width=\textwidth]{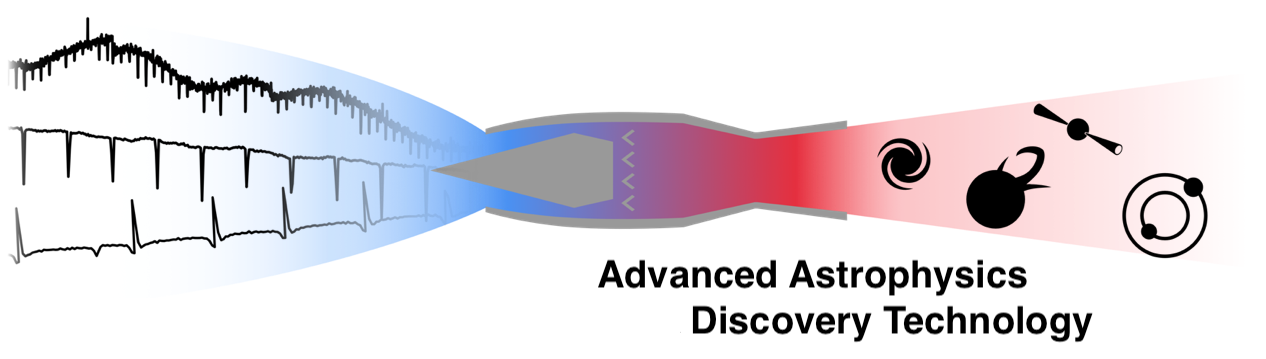}
    \label{fig:Ramjet Engine}
\end{figure}



{\large
Richard K. Barry\textsuperscript{1*},
Jogesh G. Babu\textsuperscript{2},
John G. Baker\textsuperscript{3},
Eric D. Feigelson\textsuperscript{2},
Amanpreet Kaur\textsuperscript{2},
Alan J. Kogut\textsuperscript{4},
Steven B. Kraemer\textsuperscript{5},
James P. Mason\textsuperscript{6}
Piyush Mehrotra\textsuperscript{7},
Gregory Olmschenk\textsuperscript{8}, 
Jeremy D. Schnittman\textsuperscript{3},\vspace{0.05in}\\
Amalie Stokholm\textsuperscript{9}, 
Eric R. Switzer\textsuperscript{4},
Brian A. Thomas\textsuperscript{10}, 
Raymond J. Walker\textsuperscript{11}
}
\\


\vspace*{0.3in}
{\footnotesize
\begin{enumerate}
\item Laboratory for Exoplanets and Stellar Astrophysics, NASA/GSFC, Greenbelt, MD 20771 \\
\item Dept. of Astrophysics and Statistics, Penn State University, University Park, PA 16802\\
\item Laboratory for Gravitational Astrophysics, NASA/GSFC, Greenbelt, MD 20771\\
\item Laboratory for Observational Cosmology, NASA/GSFC, Greenbelt, MD 20771\\
\item Dept. of Physics, Catholic University of America, Washington, DC 20064\\
\item Solar Physics Laboratory, NASA/GSFC, Greenbelt, MD 20771\\ 
\item Chief,  NASA Advanced Supercomputing Division, M/S 258-5, NASA Ames Research Center, Moffett Field, CA 94035\\
\item Dept. of Computer Science, City University of New York, NY 10031\\
\item Stellar Astrophysics Centre, Department of Physics and Astronomy, Aarhus University, Ny Munkegade 120, DK-8000 Aarhus C, Denmark\\
\item NASA Headquarters, Washington DC 20546\\
\item Institute of Geophysics \& Planetary Physics, University of California, Los Angeles, CA 90095-1567
\end{enumerate}
}
\vspace*{0.3in}
\emph{* Corresponding author: Richard.K.Barry@nasa.gov, 301-286-0664}\\
\vspace*{0.3in}
{\normalsize \it Submitted in response to the State of the Profession APC call; \\Decadal Survey of Astronomy and Astrophysics, 2020 }
\end{center}

\pagenumbering{gobble}

\clearpage

\pagenumbering{arabic}

\addtocontents{toc}{\vspace*{-28pt}}    
%

\addtocontents{toc}{\vspace*{5pt}}    
\vspace{-0.5em}
\section{Executive Summary}
\vspace{-0.5em}

\textbf{\underline{Motivation:}} Astrophysics is at the threshold of a new epoch in which increasingly complex, heterogeneous datasets will challenge our existing information infrastructure and traditional approaches to analysis. The rapid advancement of graphics processing units, compact field programmable gate arrays and dedicated artificial intelligence accelerator chips is now permitting the use of scientific methods, processes and algorithms to extract knowledge and insights from structured and unstructured data in ways never before seen.  Miniaturization of spacecraft architectures and supporting infrastructure is opening new observing strategies and new discovery spaces for science. The community is just beginning to awaken to these imminent challenges as evidenced by their relative lack of emphasis in the New Worlds, New Horizons ASTRO2010 decadal survey, in the ExoPAG Science Analysis Group 11 report and in the formulation of the WFIRST Data Challenge. We suggest that the Astrophysics Science Division (ASD), which has clearly recognized this new epoch of rapidly evolving information technology, could be more affirmative in its approach. We offer a modest structural solution.  
\vspace{0.5em}

\textbf{\underline{Recommendations:}}  While astrophysics, until quite recently, has existed in a relatively data-poor environment, Earth science has evolved in a data-driven domain with heterogeneous, distributed instrumentation for \textit{decades}. The Earth Science Division (ESD) provides Astrophysics with a valuable roadmap as our community faces these new challenges. We propose the creation of a new ROSES Astrophysics element, Advanced Astrophysics Discovery Technology, based loosely on ESD's Advanced Information Systems Technology element, which would seek to advance data science and new observing technology as an ongoing priority \textit{in close partnership} with the more narrowly defined science/applications elements such as ADAP, APRA and SAT.
\vspace{0.5em}

\textbf{\underline{Relevance to the ASTRO2020 Solicitation:}} Experience suggests that structural issues in how institutional Astrophysics approaches data-driven science and the development of discovery technology may be hampering the community's ability to respond effectively to this rapidly changing environment.  We stand at the confluence of a new epoch of heterogeneous multimessenger science, remote co-location of data and processing power, and new observing strategies based on miniaturized spacecraft. Significant effort will be required by the community to adapt to this rapidly evolving range of possible discovery moduses.  We offer an affirmative solution that places the visibility of discovery technologies at a level that we suggest is fully commensurate with their importance to the future of the field.  

\begin{figure}[!b]
    \centering
    \includegraphics[width=\textwidth]{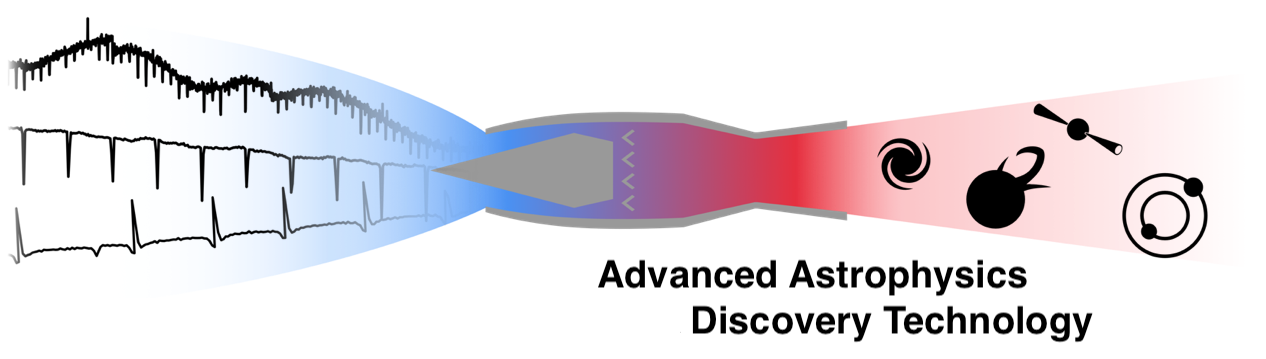}
    \label{fig:Ramjet Engine}
\end{figure}

%

\clearpage

\addtocontents{toc}{\vspace*{-3pt}}     
\section{Introduction} \label{sec:intro} 
\vspace{-.1in}
Described as a new golden age, a Renaissance or even a \textit{revolution}, there is no dispute that information technology has entered a new era at the confluence of ultra-high speed, compact processors, distributed, cloud-based\footnote{We here note that on-premises HPC compute and storage remains significantly more cost effective than commercial cloud-based solutions. See NAS Technical Report NAS-2018-01.} computing and increasingly complex, information-rich datasets.  The rapid advancement of graphics processing units (GPU), tensor processing units, compact field programmable gate arrays and dedicated artificial intelligence (AI) accelerator chips is now permitting the use of scientific methods, processes and algorithms to extract knowledge and insights from structured and unstructured data in ways never before seen. Behind the blizzard of buzzwords and seeming hype, real gains are being made in science using these new techniques.  Articles appearing on arXiv that depend on machine learning, for example, are beginning to exceed Moore's Law growth rates (Fig. 1).  Astrophysics enters this flourishing environment, well behind the learning curve, just as the available datasets become almost unmanageably large and sources of information for astrophysical phenomena proliferate in a new epoch of multi-messenger assay.  

Earth Science and, to a lesser extent, Planetary Science and Heliophysics have existed in data-rich environments for decades - migrating their evolved machine learning codes to the cloud or to institutional High Performance Computing (HPC) platforms to reside with their elaborate datasets, integrating data across multiple wavelengths and disparate instrument types and creating huge, machine-ready data archives poised for productive exploitation using state-of-the-art big data techniques. While Astrophysics\footnote{Hereafter, when capitalized, U.S. organizations charged with selecting and funding research in astrophysics. } used about 12\% of the NASA Advanced Supercomputing (NAS) Division's HPC resources in FY18 (priv. comm. P. Mehrotra, Chief, NAS Division) many practitioners in the Astrophysics Science Division (ASD) continue to work in what some have termed \textit{boutique} science in which small, rarefied, single-instrument datasets are routinely downloaded onto relatively anemic laptop computers there to be processed with traditional, ad hoc codes.  With the advent of the Large Synoptic Survey Telescope (LSST), the Transiting Exoplanet Survey Satellite (TESS) and the Wide-Field InfraRed Survey Telescope (WFIRST), these traditional methodologies will soon reach the limit of their efficacy. Indeed, many in the astrophysics community are just beginning to awaken to this challenge. The situation may be illustrated with a couple of examples, the first being WFIRST and the Science Analysis Group 11 report. Some background information is in order. 
\begin{figure}
    \centering
    \includegraphics[width=5.5in]{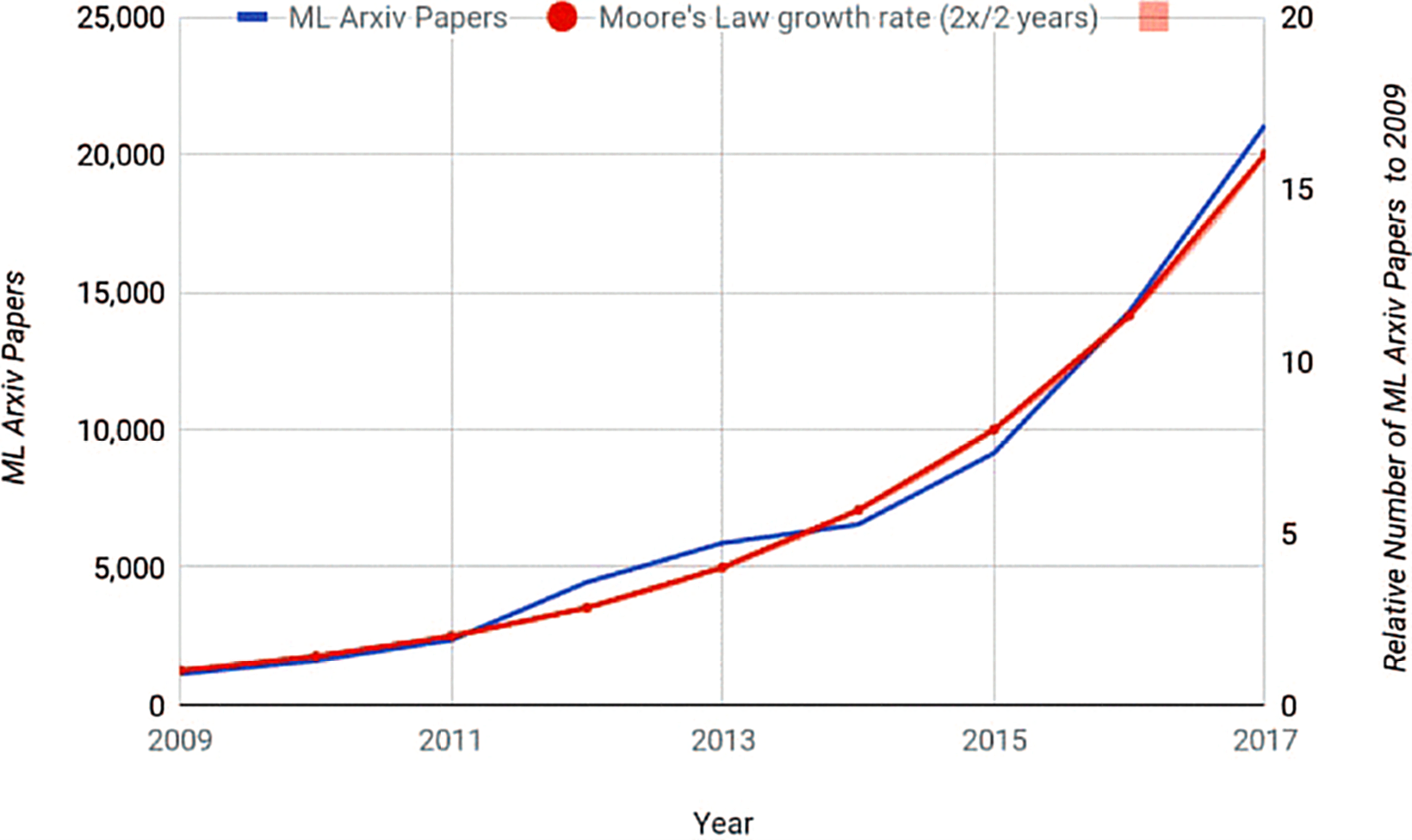}
    \caption{Machine Learning arXiv articles per year compared to Moore's Law from IEEE Micro article; Dean, Patterson \& Young, 2018}
    \label{fig:Ablation study}
\end{figure}

The 2010 Decadal Survey; New Worlds, New Horizons in Astronomy and Astrophysics \citep{Boss:2011aa}, prescribed WFIRST as its highest ranked large space project for the decade.  WFIRST, conceived as a filled aperture, wide field, near infrared (NIR) flagship observatory, was proposed to settle fundamental questions about the nature of dark energy and to complete the census of all possible types of exoplanets in the Milky Way galaxy. WFIRST's exoplanet mission \citep{Barry:2011ab} consists of two broad areas of research - gravitational microlensing and direct detection.  The first of these, microlensing, will fulfill roughly one half of the mission's core science goal by obtaining a robust sample of exoplanets from the habitable zone about their host stars out to unbound or free-floating planets. WFIRST is now in Phase-A development and is slated to launch in  late 2025.

Rather surprisingly, the technique of gravitational microlensing, upon which the WFIRST exoplanet census mission is entirely dependent, is fairly new, having detected only 31 exoplanets so far - several of which were co-discovered by principal author Barry.  The community of astronomers that specialize in this field is exceedingly insular and remarkably small - numbering perhaps 100 worldwide.  Recognizing the significant challenge this raises for the science of a NASA flagship mission, NASA's Exoplanet Exploration Program commissioned a study to be conducted by Science Analysis Group 11 (SAG11), a small team of domain scientists, to assess the current state of the field and to outline a set of concrete steps that could be taken to reduce the scientific risk and improve the scientific yield of the WFIRST exoplanet mission.  The SAG11 team met several times over a two-year period resulting in the publication of a report  \citep{Yee:2014aa} that describes the findings of the group.  The report yields three categories of programs to be pursued in preparation for the WFIRST mission, each with specific recommendations:

\begin{itemize}
    \item Programs to directly support WFIRST science and reduce its scientific risk \vspace{-0.1in}
    \item Programs that develop experience with techniques for measuring planet masses
    \vspace{-0.1in}
    \item Programs that support the development of WFIRST analysis pipelines
\end{itemize}

In response to the SAG11 report, the WFIRST project convened a small group of scientists to construct a data challenge in which many thousands of simulated, classified, microlensing lightcurves were produced and made available to the broader science community.  This was undertaken to spur interest in microlensing modeling efforts with an open-source code developed for this express purpose.  

Close scrutiny of the SAG11 report and the goals of the data challenge reveal that there was little recognition of the enormity of the task of actually \textit{locating} planetary microlensing events in full frame WFIRST images.  This oversight is significant given that the WFIRST exoplanet microlensing mission will tile the dense star fields of the Galactic bulge with an unprecedentedly massive array of detectors, totalling over 300 million pixels, at a rate of one integration every 15 minutes. The SAG11 team, \textit{including R.K.Barry, the corresponding author of this ASTRO2020 submission}, convened specifically to identify technical and scientific challenges to be met by WFIRST, clearly missed this significant exigency.  The WFIRST data challenge team, likewise, did not identify recovery of microlensing signals in full frame images as a significant obstacle.  This suggests that at least some in the astrophysics community may be unprepared for the manifest difficulty of simply finding specific astronomical signals in such a torrent of data.  Advanced information systems technology in general and machine learning techniques in particular will be essential to assist them, allowing researchers to focus their efforts on the science rather than on the search for relevant information.  There is, however, surprisingly little activity in the astrophysics community on this issue.

A second example may be found in the community's approach to ground observations in gravitational microlensing. These have been conducted primarily by the Optical Gravitational Microlensing Experiment (OGLE) and the Microlensing Observations in Astronomy (MOA) observatories.  An explanation of how MOA microlensing observations are presently conducted is in order to illustrate how advanced information technology could facilitate the science of such a survey observatory.  During the season when the Galactic bulge is near zenith in the Southern hemisphere where MOA is located, teams of two usually operate the observatory.  Typically, the observing team may include an experienced observer and one or possibly two graduate students.  Once the observatory is cooled down and prepared for the night's observations, most of the observatory function is conducted under computer control.  As the telescope is driven to tile the MOA field of regard (FOR), images come in, are differenced and any source found to have changed in brightness from a deep reference image is presented, one at a time, to the observer.  The observer, through experience, must make a determination of the nature of the growing lightcurve by eye and press one or more buttons to flag it or discard it.  The overwhelming majority of changes identified by differencing with the reference frame are not due to microlensing.  Indeed, most are due to the movement of dust on the detector or near-field moving objects.  A large number are due to other sources of variability such as binarity, flares, etc. The observer flags any change that appears to be of significance and that change is tracked from one exposure to the next - in turn with other candidate objects.   The Unix cluster at the observatory generates simple point source, point lens (PSPL) models in an attempt to fit such flagged objects.  If there is a strong trend towards a fit, the observer flags the object as a candidate microlensing event.  If the event is fit by the PSPL model and appears to be of high magnification - an event with the highest probability of exoplanet detection because the source star images are large - a bulletin is sent out to the consortium and a number of small, narrow-field instruments are trained on the object to obtain observations at high cadence and with broader phase coverage.  All sources of variability that do not appear to be microlensing are simply discarded.  As should be obvious to the reader, much is lost in this quite tedious process.  

Current machine learning techniques could be employed for both of these classification tasks - MOA and WFIRST -  with the potential to greatly increase the science yield of the observations, not just in gravitational microlensing, but across the board in time domain astronomy.  There is, however, no obvious mechanism to fund such data science work in which the proximate science case may be less well-formed.  Indeed, work that is focused on the development of data-driven discovery tools \textit{may have no singular instantiating science case at all}. Clearly, such work is increasingly urgent, but frustratingly difficult to find direct support for. We argue that this may largely be due to structural issues in how Astrophysics approaches funding research into data driven processes.  We expand on this idea below.

\addtocontents{toc}{\vspace*{-3pt}}     
\vspace{-.2in}
\section{Information Challenges in a New Epoch} \label{sec:problem} 

\vspace{-.1in}
\subsection{Generalization of the Problem}
In broadest terms, Astrophysics, in our considered opinion, has not been as responsive to the broader issues associated with our transition to a data driven regime as we would wish.  Rather than vigorously encouraging active research into advanced information technology in the service of astrophysics, Astrophysics appears to be just beginning to apprehend the true urgency of the challenge we face. Indeed, many in the astrophysics community appear ever ready to fall back on a boutique science approach, an approach that fits well with, and is certainly not discouraged by, the current structure of Astrophysics. Where monotonous classification tasks are required, for example, we frequently see that the traditional utilization of graduate students (or, in a modern incarnation, \textit{Citizen Scientists}) is instinctively prescribed as a solution.\footnote{We here note that state-of-the-art ML techniques have begun outperforming humans in many pattern recognition tasks \citep{2017arXiv170606969G}.}  Researchers in astrophysics cannot and should not, however, be accused of intellectual laziness in this regard.

Advanced information moduses such as AI and ML have \textit{steep} learning curves, requiring much study and experience to master.  The field is in extremely rapid flux with new technologies emerging at a pace that is challenging to match for long enough to develop a workable solution to an astrophysical question. For example, Theano, a deep learning python library and compiler optimized for high-speed matrix mathematics on CPUs and GPUs was, until recently, the go-to design space for astrophysics ML and AI applications.  As of 2017, Theano is no longer supported and has been almost completely supplanted by  other options including TensorFlow, an open source AI framework associated with Google, and CUDA,  a collection of NVIDIA libraries, tools, compilers and application programming interfaces.  An astronomer hoping to create an application based on these tools would be right to be wary to choose one and launch into months of self-teaching and development, however, as these will almost certainly be supplanted in the very near future by such suites of tools as NVIDIA's RAPIDS - a completely re-imagined environment for the creation of end-to-end data science and analytics pipelines entirely on GPUs. Which environment should an established astrophysicist invest time in learning? Will it solve the scientific problem at hand?  Will that development architecture even exist in five years when the observatory is ready for science operations?   

Software is not the only issue.  New observing strategies are emerging in response to the rapid evolution in very capable miniaturized hardware.  Graphics processing units, tensor processing units, compact field programmable gate arrays and dedicated AI accelerator chips are under development and are being radiation and vacuum tested for space-borne applications potentially increasing the science return capabilities of SmallSat and CubeSat architectures.  For example, some new CubeSat designs would leverage new fanless GPU processors or dedicated AI accelerators to triage data as it is accumulated, downlinking only the germane scientific information when sufficient communications power is available, thus opening up the possibility of much more distant orbits for modest-sized spacecraft.  How much power is needed to conduct preprocessing on the AI accelerator?  How will that affect the overall power budget of the spacecraft?  Would it be better to simply downlink full frame images and do the processing on the ground?  Should I choose an ASIC instead of an FPGA?  What about compressive sensing? Will that affect my observing strategy? 

A great deal of domain-specific expertise is needed to harness these rapidly evolving technologies for both information systems and for observing strategies.  These discovery technologies are at the nexus of scientific inquiry and cutting-edge information technology.  Yet they are more of an apt means to open new scientific discovery spaces rather than answer a specific scientific question. Discovery technologies are different in an essential way than other moduses that fit more naturally into the present Astrophysics funding framework - \textit{discovery technologies often lack a singular instantiating science case.} 

\vspace{-.15in}
\subsection{Programmatic Structure Considerations}
\vspace{-0.05in}
Presently, in the case of NASA, its Research Opportunities in Space and Earth Sciences (ROSES) program offers several Astrophysics program elements. We will describe the goals of each of these in broad terms and show why they may not fit the discovery technology moduses we described previously. 

The ROSES Astrophysics Data Analysis Program (ADAP) supports research with an emphasis primarily on the analysis of archival data from NASA missions - both current and past. While the ADAP element does encourage studies that may include data from several observatories and/or wavelength ranges, ADAP typically supports studies that are based on and in response to one specific scientific question that may be answered with the proposed strategy. This is not a natural fit for a discovery technology proposal.  For example, a proposal to assess several machine learning approaches to the efficient classification of photometric lightcurves would be unlikely to find support within the ADAP framework as it is a proposal to develop a discovery mechanism that has no singular science goal.  Naturally, the proposers \textit{could} describe a motivating science objective and propose a machine learning approach to locating the lightcurves associated with that particular objective, but the assessment of different ML approaches to the problem in search of the most efficient methodology would be considered by the typical jury as a significantly weakening aspect - unjustified by the scientific question.  Within ADAP, the data processing mechanisms proposed must be tightly linked to a particular instantiating science question.  This simply is not the case for the development of a discovery technology. Indeed, we have found through experience in the writing of a large number of proposals to the ADAP element that the data analysis component of a proposal must be a direct, minimalist and closely proximate solution to a well-formed, tightly defined science question or support is unlikely to be attracted.  A broader approach to opening a discovery space is typically dismissed as mere data science `adventurism' by reviewers for this Astrophysics element especially if the postulated scientific question is perceived to be of insufficient gravity to justify a more thorough development effort - even when that effort may result in a number of additional findings or the creation of a broadly useful discovery tool.

The ROSES Astrophysics Research and Analysis Program (APRA) supports suborbital investigations, development of detectors and supporting technology, laboratory astrophysics, and ground based observing.  The supporting technology component of APRA supports the development of new data analysis methods that are directly applicable to future space missions.  While the development of discovery technology such as broadly applicable machine learning codes \textit{could} fit into this subcategory of APRA, any such proposal submitted in response would, again, need to directly compete with investigations targeted at, and predicated on, expeditiously delivering the answer to a singular instantiating science question or closely associated set of questions.  As in the ADAP element, this puts discovery technologies that may require significant investment in time yet with an unclear association to any specific science question at a significant disadvantage. Our own experience in writing proposals to this element strongly supports this perception. 

The ROSES Astrophysics Theory Program (ATP) supports theoretical investigations or modeling of the astrophysical phenomena targeted by past, current, or future NASA astrophysics space missions.  ATP is \textit{quite narrowly} focused on specific science questions and the mechanisms to achieve answers to those questions and would not be a mechanism for general discovery technologies.  Strategic Astrophysics Technology (SAT) would seem, from its name alone, to be a good fit for proposals targeting discovery moduses.  SAT, however, is primarily targeted at advancing the  readiness level of  technologies addressed to and meant to be directly infused into future spaceflight missions. Proposals to develop discovery moduses meant to utilize cutting-edge data technologies to open discovery spaces for the astrophysics community would be again at a relative disadvantage.  The need for immediate, thoroughgoing change in our approach to the data would be, as in APRA and ADAP, `lumped in' with proposals that may depend on solutions promising a quick payout with a traditional approach. We argue below that discovery moduses should be considered separately due to the significantly different expertise needed in execution and in the formulation of juries that are tasked with assessment of such proposals.

\addtocontents{toc}{\vspace*{-3pt}}     
\vspace{-.2in}
\section{A Proposed Structural Response} \label{sec:solution} 
\vspace{-.1in}

We suggest that the current structure of Astrophysics could be more responsive to this new epoch of rapidly evolving information technology.  As noted in Section \ref{sec:intro}, Earth Science has existed in a data-rich environment for decades. Indeed, the National Academy of Sciences' Decadal Strategy for Earth Observations from Space (NAS 2018) contains specific recommendations to NASA that recognizes the importance of data-driven technologies.  Specifically:
\begin{itemize}
    \item Recommendation 4.2: "Success in observation-driven modeling holds the key for maintaining end-to-end capability..."
    \vspace{-0.1in}
    \item Recommendation 4.3: "NASA, NOAA and USGS should continue to advance data science as an ongoing priority within their organizations in partnership with the science/applications communities"
\end{itemize}

The Survey also gives specific recommendations for the optimization of data archives for machine learning and AI readiness:

\begin{itemize}
    \item Recommendation 3.3.1: "NASA should work toward Analytics-Optimized Data Stores for data to serve as building blocks for analytic tools and services, with a focus on building and exposing APIs..."
    \vspace{-0.1in}
    
    \item Recommendation 3.5.1: "NASA HQ should provide support to R\&D efforts that evaluate new strategies to create labeled training data in Earth Science..."
    \vspace{-0.1in}
    
    \item Recommendation 3.5.2 "NASA HQ should develop an approach to systematically create, distribute and archive training/labeled data..."
\end{itemize}

Due to its data-rich environment, The Earth Science Division (ESD) of NASA has long been responsive to changes in data science with a programmatic structure that reflects the priority it places on discovery technologies.  In particular, we note that Earth Science has an \textit{entire program element}, Advanced Information Systems Technology (AIST), dedicated to this modus.  The ESD further breaks down the element into two logical domains: New Observing Strategy (NOS) and Analytic Center Framework (ACF).  The particular formulation of these two domains gives insight into how the AIST contributes to the mission of the ESD. 

The ACF "...integrates new or previously unlinked datasets, tools, models, and a variety of computing resources together into a common platform to address previously intractable scientific questions. Additionally, this activity seeks to generalize custom or unique tools that are used by a limited community, in order to make them accessible and useful to a broader community." This domain supports activities and products such as:

\begin{itemize}
    \item Data-driven modeling tools enabling the forecast of future behavior of the phenomena
    \vspace{-0.1in}
    \item Innovative visualization, including virtual and augmented reality, to enable scientific understanding
    \vspace{-0.1in}
    \item Ingesting data from various sources into a temporary storage system and development of a publishable description of the data
\end{itemize}

The NOS domain "... provides a framework for identifying technology advances needed to exploit newly available observational capabilities. The NOS thrust enables development of the information technologies needed to support planning, evaluating, implementing, and operating a dynamic set of observing assets consisting of various instruments located at different vantage points (e.g., in situ, airborne, and in orbit) to create a more complete picture of a natural phenomenon or physical process."  A few illustrative domain activities include:

\begin{itemize}
    \item Evaluation/comparison of alternative observing strategies
    \vspace{-0.1in}
    
    \item Integrated operations of different types of instruments or at different vantage points
    \vspace{-0.1in}
    
    \item Intercalibration of heterogeneous instruments
\end{itemize}

It would be instructive to examine how such a structure might benefit a few Astrophysics domain projects selected from  science white papers submitted to the ASTRO2020 survey.

\vspace{-.15in}
\subsection{Example Enabled Domain Projects}
There are a large number of science activities submitted this Spring to ASTRO2020 that would be well-served by an AIST-type Astrophysics element. Of 33 white papers describing new surveys, almost all of them will benefit directly from generalized machine learned classification approaches and accumulation and post-processing of data to make it machine ready.  For example, ASTRO2020 Response 346 describes the potential to measure the Hubble constant from gravitational wave standard sirens to 1-2\% uncertainty. This will require a new generation of spectroscopic galaxy surveys to achieve such precision.  Such surveys will likely need significant new data infrastructure and will almost certainly require the use of new machine learning procedures to identify galaxies associated with the GW sirens. 

Four white papers directly address the need to advance data analytics infrastructure.  For example, ASTRO2020 Response Number 482 describes Cyberinfrastructure Requirements to Enhance Multi-messenger Astrophysics in which they argue for a separate \textit{institute} to meet the challenges of the rapidly burgeoning heterogeneous astrophysics data immanently impinging on the field. Five white papers deal with making data machine learning ready and co-locating it with HPC or cloud-based compute resources.  For example, ASTRO2020 Response 55 makes the case for Euclid/LSST/WFIRST Joint Survey Processing that will require transformation of all three data sets into a common architecture. There are a large number of other white papers that would fit well into an AIST-like structure.  For example, Three white papers submitted to ASTRO2020 directly address the development of convolutional neural nets for two and three dimensional data from heterogeneous instruments. There are several dozen that would fit neatly under the NOS domain and we expect this number to rise sharply once white papers submitted to the ASTRO2020 APC call are included. 

While scientific projects such as those described above \textit{could} be made to fit into present Astrophysics elements, an AIST-like element would serve as a more apt vehicle for them and could motivate the congress of evaluation juries that embody expertise peculiar to data analytics, machine learning and new discovery techniques. We suggest that expertise and work in these rarefied domains may be insufficiently supported and encouraged by the current Astrophysics elemental structure. Given the rapid increase we perceive in quantity and source heterogeneity of astrophysics data and in nascent observing strategies enabled by a trend towards instrument and spacecraft miniaturization, a structural change in Astrophysics seems justified. 
\vspace{-0.2in}
\subsection{Towards A New Program Element}
 We have outlined what we see as a significant challenge faced by Astrophysics in the coming decade. We perceive multiple changes to the very nature of our work all likely to occur within the coming generation.  Cutting edge discovery technologies will require the deliberate cultivation of a new set of skills in the astrophysics community and will need to be carefully encouraged and nurtured by a structure that places the visibility and import of these new moduses at a level commensurate with their importance to the future of our field. The old model of single wavelength range, single instrument boutique science is unlikely to carry us much further - although there is almost palpable reluctance to abandon these traditional approaches.  There is, for example, great reluctance in the astrophysics community to co-locating data and HPC or cloud-based compute power. There is much reluctance to and little support for revisiting legacy datasets to make them machine/AI ready. These retrograde tropes, and others we could describe, if permitted to continue into this new data-driven era, will almost certainly impede scientific progress. 
 
 We suggest that the best way to guard against disadvantageous institutional inertia is to change the institutional structure at a fundamental level through the inclusion of a new Astrophysics program element. This new focus, which we tentatively name the Advanced Astrophysics Discovery Technology (AADT) element, would seek to advance data science as an ongoing priority \textit{in close partnership} with the more narrowly defined science/applications elements (e.g. SAT, APRA, ADAP). Much may be learned from the experience of the Earth Science Division and we strongly recommend obtaining the advice of the ESD's management in the formulation of the AADT. This new element will almost certainly be different from the AIST, but Astrophysics would clearly benefit from the ESD's long experience in balancing support for data-driven discovery technologies with more narrowly defined science elements.

The examples we have given to motivate consideration of the AADT and the arguments founded on those examples are not dispositive.  We have not shown that discovery technology couldn't simply fit within the current framework.  However, we submit that a new program element is justified and would bring emphasis and visibility where it is now lacking.  We could simply wait for generational change to bring the needed skills into astrophysics over time or we may choose to decisively act. \textit{We recommend a well-considered, modest yet affirmative change to the elemental architecture of Astrophysics in response to this known imminent challenge.} 

We are entering an era in which it would not be an exaggeration to say that there is a seismic shift in the quantity and complexity of astrophysics science data and in the technologies to obtain that data. Significant effort will be required by the community to adapt to a rapidly evolving range of possible discovery moduses.  Indeed, in a broad range of cases, traditional methods will need to be simply discarded and radically new methods embraced. To relegate support for such efforts as a mere subcategory of an sub-element in NASA's ROSES portfolio would, in our opinion, give them insufficient emphasis and exposure. Once buried in a structure targeted mainly at a traditional approach to astrophysics, new discovery technology moduses would be unlikely to establish support commensurate with the importance and immediacy of the need to develop these capabilities.  Moreover, the imperative for Astrophysics to cultivate the rarefied expertise required by such advanced astrophysics discovery technologies and to be in a position to stand up committees with appropriate expertise to knowledgeably assess AADT-based proposals, is unlikely to be sufficiently supported by the existing structure.

\begin{center}
  $\ast$~$\ast$~$\ast$
\end{center}

We are at the threshold of a new epoch in astrophysics, a \textit{data-driven} epoch in which information from disparate sources and heterogeneous instruments must be combined in new ways and in which subtle signals imprinted on vast data troves at the very largest scales must be detected to advance our science.  This is an especially important time - at the confluence of much higher data rates, multi-messenger assay, miniaturization of observatory components and a burgeoning machine learning and data analytics field. This is the moment to make fundamental changes in our approach to astrophysics.  We are entering the decade of Advanced Astrophysics Discovery Technology.

\clearpage


\addtocontents{toc}{\vspace*{-16pt}}	
\addcontentsline{toc}{part}{\normalsize \emph{References and Acronyms}}
\addtocontents{toc}{\vspace*{-8pt}}	

\renewcommand{\refname}{}
{\Large \noindent \emph{References and Acronyms}} \\[-10ex]
\renewcommand{\bibsep}{5pt}	
\bibliographystyle{chicago}
\bibliography{bibliography.bib}

\clearpage

\label{sub:acronyms}

\begin{table}
\caption{Table of acronyms.}
\label{tab:Acronyms}
\centering
\begin{tabular}{|>{\raggedright}p{0.22\textwidth}|>{\raggedright}p{0.22\textwidth}||>{\raggedright}p{0.22\textwidth}|>{\raggedright}p{0.22\textwidth}|}
\hline
ASD & Astrophysics Science Division & ESD & Earth Science Division \tabularnewline
\hline
HPC & High Performance Computing &FFI&Full Frame Image
\tabularnewline
\hline
GPU&Graphics Processing Unit&NAS&NASA Advanced Supercomputing Division
\tabularnewline
\hline
LSST&Large Synoptic Survey Telescope&TESS& Transiting Exoplanet Survey Satellite
\tabularnewline 
\hline
ML&Machine Learning&MOA&Microlensing Observations in Astrophysics
\tabularnewline
\hline
NASA&National Aeronautics and Space Administration&SAG11&Science Analysis Group 11

\tabularnewline
\hline
ExoPAG&Exoplanet Exploration Program&FOR&Field of Regard
\tabularnewline
\hline
PSPL&Point Source Point Lens&FPGA&Field Programmable Gate Array
\tabularnewline
\hline
ADAP&Astrophysics Data Analysis Program&APRA&Astrophysics Research and Analysis Program
\tabularnewline
\hline
WFIRST&Wide-Field InfraRed Survey Telescope& ATP & Astrophysics Theory Program
\tabularnewline
\hline
AADT&Advanced Astrophysics Discovery Technology&AIST&Advanced Information Systems Technology 
\tabularnewline
\hline
%

\end{tabular}
\end{table}

\end{document}